# Interpretation of wake instability at slip line in rotating detonation

Pengxin Liu[a], Qin Li[a, b], Zhangfeng Huang[c], Hanxin Zhang[a, b]

[a]State Key Laboratory of Aerodynamics, Mianyang, Sichuan, 621000, China
[b]National Laboratory of Computational Fluid Dynamics, Beijing, 100191, China
[c]TianJin University, Tianjin, 300000, China

**Abstract**

In studies on instabilities of flowfield in rotating detonation, one of the most common concerns is the instability at the slip line originating from the conjunction of the detonation wave and oblique shock. Using Euler equations associated with 7-species-and-8-reaction finite-rate chemical reaction model of hydrogen/air mixtures, further studies are performed to simulate the 2-D rotating detonation, and the flow mechanism of instability at the slip line is investigated in depth. The results show that the distinct wake profile exists at the slip line, which is different from the typical mixing layer. Analysis indicates that the generation of wake is caused by the transition shock between the detonation wave and oblique shock. Because of the wake profile, the vorticity distribution therein appears in a double-layer layout, and different evolutions exist in different vorticity layers. Based on the velocity profile across the slip line, the analysis by the linear stability theory is made, and two unstable modes which have different shape profiles and phase velocities are found. Discrete Fourier transformation is utilized to analyze the numerical results, and similar shape profiles are obtained. A general coincidence in velocity of vortex movement is also attained between the theoretical predictions and simulations. Investigations show that the wake instability is responsible for the unstable mechanism, and corresponding unstable structures differs from the canonical ones in typical mixing layers.

**Key words:** Rotating detonation; Slip line; Kelvin-Helmholtz instability; Wake instability

## 1. Introduction

As the shock-induced combustion, detonation theoretically has advantages over traditional deflagration on high thermal efficiency and fast reaction rate. Several types of detonation engine have been proposed as propulsion power, among which rotating detonation engine (RDE) has received wide attention in recent years. Compared with other types, RDE needs a single time start-up and has a relative stable thrust, which seems to be more attractive for engineering.

Investigations on the rotating detonation have been conducted experimentally and theoretically since 1960s by Nicholls et al.[1] and Voitesekhovskii[2]. In the next few decades, topics regarding the performance of RDE have been studied [3-7], e.g., ignition methods, injection schemes, compositions of fuel and oxidant, configurations of combustor, operation window. Usually in RDE, rotating detonation waves run in a solid annular chamber at the speed of kilometers per second, which is difficult for experiments to explore flow subtleties. In this regard, computation can serve as a convenient and cost-saving tool for investigation.

With the advancement of CFD, achievements have been attained on concerns of engineering and fundamental researches, such as thrust performance [8, 9], thermodynamic analysis [10], design of chamber configurations [11], three-dimensional effect [12], detonation wave structures [13], the stability of detonation [14].

Because of the possible influence on RDE operation, flow instabilities are concerned by engineering. Among the unstable complexities, the instability at the slip line originating from the conjunction between the detonation wave and oblique shock is the most typical one referred by literatures [8-17]. On the two sides of the slip line, the gas states are different, and especially velocity difference is assumed to exist. Such situation is apt to yield instabilities, and vortex-like structures are usually observed along the slip line. Many investigations have reported such unstable structures [8-9, 11-17], and it is generally believed that Kelvin-Helmholtz (K-H) instability is responsible for their generation [15-17]. Although the wide reference of the phenomenon, detailed explanation and thorough analysis are absent so far according to the authors' knowledge. It is well-known that K-H instability is regarded as the mechanism for the evolution of mixing layers, which is characterized by two stratified flows with different velocities. In reality, the velocity discontinuity is often smoothed by connections therein like hyperbolic tangential profile. In such situation, one single layer of vorticity exists and its loss of stability yields roll-up of vortices. In the case of rotating detonation, similar vortex-like structures occur along the slip line, but other additional ones have been observed also. In Refs. [11, 13], Schwer et al. clearly manifested that two layers of unstable structures existed along the slip line, i.e. besides the primary vortex-like structure, a secondary instability in wavy form was observed superimposing on the former (e.g., see Fig. 2 in Ref. [11]). Such phenomenon differs from that in canonical mixing layers, therefore it is wondered that more complexities exist in the slip line of RDE.

Targeting at above uncertainties, further investigations are carried out on the flow mechanism of instability at the slip line, and an understanding of the wake instability is proposed which is different from the situations in canonical mixing layers. Prior to detailed discussions, computation methods and corresponding validations are introduced in section 2. Then the overall features of rotating detonation are discussed in section 3. The wake instability at the slip line is proposed and analyzed in section 4. A brief conclusion is drawn in section 5.

## 2. Physical model and numerical method

### 2.1. Physical model

The typical combustion chamber in RDE research consists of two coaxial cylinders. Because the radial width of the chamber is usually quite small compared to the azimuthal and axial length-scales, the radial effect can be ignored and the chamber is usually simplified into 2-D one for investigation as shown in Fig. 1. In the figure, fuel and oxidizer are injected into the chamber through slits or orifices at the left end, and one or several triangle layers of fresh mixture are formed thereafter. The height of mixtures keeps increasing until one side of the layer meets the detonation wave and the combustion happens. The process is re-circulating circumferentially around the combustion chamber with an oblique shock connecting to the detonation wave. Because the states of products after the detonation wave are different from those after the shock wave, a slip line which is referred as by most literatures is generated.

The combustion products are exhausted at the right open-end. In Fig. 1, *x*-axis corresponds to the axial direction, and *y*-axis is related to azimuthal direction.

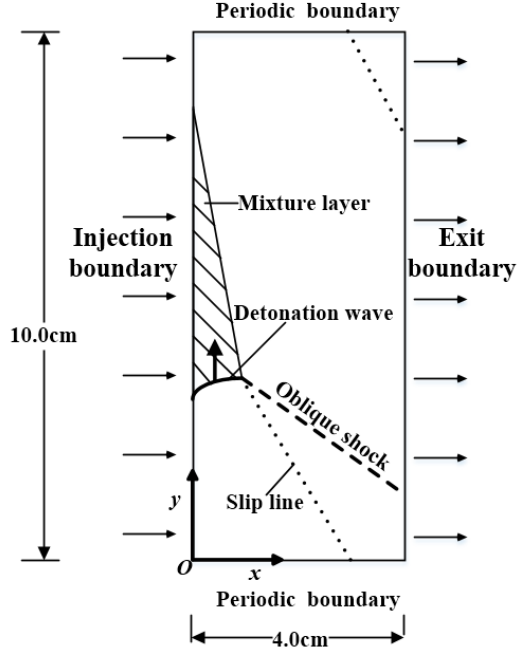

Fig. 1. Illustration of 2-D simplification of RDE.

## 2.2. Governing equations

As usually adopted in literatures [8-17], contributions of viscosity, thermal conduction and mass diffusion are often ignored for simplification, and therefore the compressible Euler equations with chemical source terms are chosen as governing equations. For the sake of applicability, 3-D equations are chosen to solve 2-D problem. The vector form of equations in Cartesian coordinates $(x, y, z)$ are given as

$$\frac{\partial Q}{\partial t} + \frac{\partial F}{\partial x} + \frac{\partial G}{\partial y} + \frac{\partial H}{\partial z} = S, \tag{1}$$

where $Q = \begin{pmatrix} \rho \\ \rho u \\ \rho v \\ \rho w \\ E \\ \rho f_i \end{pmatrix}$, $F = \begin{pmatrix} \rho u \\ \rho u^2 + P \\ \rho uv \\ \rho uw \\ (E+P)u \\ \rho u f_i \end{pmatrix}$, $G = \begin{pmatrix} \rho v \\ \rho vu \\ \rho v^2 + P \\ \rho vw \\ (E+P)v \\ \rho v f_i \end{pmatrix}$, $H = \begin{pmatrix} \rho w \\ \rho wu \\ \rho wv \\ \rho w^2 + P \\ (E+P)w \\ \rho w f_i \end{pmatrix}$, $S = J^{-1} \begin{pmatrix} 0 \\ 0 \\ 0 \\ 0 \\ 0 \\ S_i \end{pmatrix}$.

Further explanations about related variables are given as follows. $\rho$ and $P$ are the density and pressure of the mixture. $u$, $v$ and $w$ are velocity components. $f_i$ is the mass fraction of the *i*-th component (*i*=1…*Ns*-1 and *Ns* is the number of species), The total energy $E$ is defined as:

$$E = \rho \sum_i f_i h_i - p + \frac{1}{2}\rho(u^2 + v^2 + w^2), \qquad (2)$$

where $h_i$ is the specific enthalpy. $h_i$ is derived by using polynomial fitting of the temperature, and coefficients in the polynomial can be found in JANAF table [20]. $S_i$ is the mass production rate of the $i$-th species, which is derived according to Ref. [12] as

$$S_i = M_i \sum_{j=1}^{Nr}(\beta_{ji} - \alpha_{ji})\left( K_j \prod_{i=1}^{Ns}\left(\frac{\rho_i}{M_i}\right)^{\alpha_{ji}}\left(\sum_{k=1}^{Ns}\frac{\rho_k}{M_k}C_{jk}\right)^{L_j} - K_{-j}\prod_{i=1}^{Ns}\left(\frac{\rho_i}{M_i}\right)^{\beta_{ji}}\left(\sum_{k=1}^{Ns}\frac{\rho_k}{M_k}C_{jk}\right)^{L_j}\right), \qquad (3)$$

where $M_i$ is the molecular weight, $\alpha_{ji}$ and $\beta_{ji}$ are stoichiometric coefficients of the $i$-th species in the $j$-th reversible reaction equation, $Nr$ represents the number of reactions, $L_j$ represents whether the $j$-th reaction includes triple-object collision with the value "0" for No or "1" for Yes, $C_{jk}$ is the collision efficiency of the $k$-th species in the $j$-th reaction and is set as "1" in this computation, and $K_{fj}$ and $K_{bj}$ are the reaction rates in forward and reverse direction. If the reversible reaction equation has the form as $\sum_{k=1}^{Ns}\alpha_{jk}X_k \Leftrightarrow \sum_{k=1}^{Ns}\beta_{jk}X_k$ where $j=1, ..., N_r$ and $X_k$ denotes the species such as $O_2$, $K_j$ and $K_{-j}$ are evaluated by Arrhenius Law as

$$\begin{cases} K_j = A_j T^{B_j} e^{\left(-C_j/T\right)} \\ K_{-j} = A_{-j} T^{B_{-j}} e^{\left(-C_{-j}/T\right)} \end{cases}, \qquad (4)$$

where $A_j, B_j, C_j$ and their reversible counterparts are species-related coefficients. For brevity, further details are omitted and suggested to Ref. [25]. Through Eqs. (1)-(4) and $p = \rho RT$ with $R = R_0 \sum_{i=1}^{Nr}\frac{f_i}{M_i}$ ($R_0$ is the universal gas constant), the general $Ns$-species-and-$Nr$-reaction model can be integrated with aerodynamic equations. In current study, a specific 7-species-and-8-elementary reaction model of hydrogen/air mixtures (7 species: $H_2$, $O_2$, H, O, OH, $H_2O$, $N_2$) is adopted as that in Ref. [21].

**2.3. Numerical methods**

In order to achieve high resolutions of the subtle flow structures and maintain the numerical stability in the meanwhile, a fifth-order WENO-type scheme, namely WENO-PPM5[23], is applied to discretize the inviscid flux $F$, $G$ and $H$. The reason to use

WENO-PPM5 other than the canonical fifth-order WENO scheme is that the former has manifested higher resolution to describe subtle flow structures along the slip line[23], which is just needed in current investigation. A brief review of the scheme is given as following.

Taking the 1-D hyperbolic conservative law as an example,

$$u_t + f(u)_x = 0. \tag{5}$$

Supposing grids are uniform with the interval $\Delta x$, Eq. (5) at $x_j$ can be re-written as

$$(u_t)_j = -\left(\hat{f}_{j+1/2} - \hat{f}_{j-1/2}\right)\big/\Delta x, \tag{6}$$

where $\hat{f}_{j+1/2}$ is the evaluation of $\hat{f}(x)$ at $x_{j+1/2}$, and $\hat{f}(x)$ is implicitly defined by $f(x) = \frac{1}{\Delta x}\int_{x-\Delta x/2}^{x+\Delta x/2} \hat{f}(x')dx'$. Usually, the flux $f(x)$ is split into the positive part $f^+$ and negative part $f^-$ according to eigenvalues of $\partial f(u)/\partial u$, and so does $\hat{f}(x)$ accordingly. Taking $f^+$ as an example and dropping the superscript '+' for brevity, the standard WENO5 scheme can be formulated as [22]:

$$\hat{f}_{j+1/2} = \sum_{k=0}^{2} \omega_k q_k, \tag{7}$$

where $\omega_k$ is the nonlinear weight, $q_k$ denotes the candidate scheme on basic stencil by $q_k = \sum_{l=0}^{2} a_{k,l} f\left(u_{j+k+l-2}\right)$, and $a_{k,l}$ can be found in Ref. [22]. $\omega_k$ is obtained by

$$\omega_k = \alpha_k \bigg/ \sum_{l=0}^{2} \alpha_l \quad \text{and} \quad \alpha_k = C_k \big/ (\varepsilon + IS_k)^2, \tag{8}$$

where $C_k = \{0.3, 0.6, 0.1\}$ and usually $\varepsilon = 10^{-5} \sim 10^{-7}$. $IS_k$ is the smoothness indicator and the one in Ref. [22] has the form

$$\begin{cases} IS_0 = \frac{13}{12}\left(f_{j-2} - 2f_{j-1} + f_j\right)^2 + \frac{1}{4}\left(f_{j-2} - 4f_{j-1} + 3f_j\right)^2 \\ IS_1 = \frac{13}{12}\left(f_{j-1} - 2f_j + f_{j+1}\right)^2 + \frac{1}{4}\left(f_{j-1} - f_{j+1}\right)^2 \\ IS_2 = \frac{13}{12}\left(f_j - 2f_{j+1} + f_{j+2}\right)^2 + \frac{1}{4}\left(3f_j - 4f_{j+1} + f_{j+2}\right)^2 \end{cases}. \tag{9}$$

In Ref. [23], new piecewise-polynomial mapping functions $g_k$ were proposed to improve the performance of WENO5, which was fulfilled by invoking a revision on $\omega_k$

obtained by Eq. (8). A fifth-order version of mapping functions is used here as

$$\omega'_k = g_k(\omega_k) = \begin{cases} C_k\left[1+(a-1)^5\right] & \text{if } \omega_k \leq C_k \\ C_k + b^4(\omega_k - C_k)^5 & \text{otherwise} \end{cases}, \quad (10)$$

where $a = \omega_k/C_k$ and $b = 1/(C_k - 1)$. At last, the final nonlinear weights $\omega_k$ will be acquired by normalizing the newly-obtained $\omega'_k$ through $\omega_k = \omega'_k / \sum_{l=0}^{2} \omega'_l$. The corresponding scheme is called as WENO-PPM5. More details are suggested to Ref. [23]. The scheme for the negative part of flux can be derived according to the symmetry of the algorithm with respect to $x_{j+1/2}$.

For the temporal derivative in Eq. (1), a third-order TVD Runge-Kutta scheme [22] is used for discretization. For the flux splitting referred above, Steger-Warming splitting method is used in current study.

### 2.4. Boundary conditions

As shown in Fig. 1, three kinds of boundaries are involved in this study, i.e., injection boundary, periodic boundary and exit boundary. In real injection system, the fuel and oxidant are fed into the combustion chamber through slits or orifices separately. To simplify the complexity such as mixing process and injection specifics, numerical models are usually employed [8-18, 24]. One commonly-used assumption is that the fuel and oxidant are pre-mixed at stoichiometric ratio and are injected continuously into the chamber from the left end. In this study, the treatment of Ref. [24] is chosen. Specifically, the injection can be classified into three situations depending on the left-wall pressure $p_w$ which is extrapolated from the interior field:

(1) For $p_w \geq p_0$ where $p_0$ denotes the total pressure, no injection happens. All variables are extrapolated from the interior field except that the axial velocity $u = 0$.

(2) For $p_{cr} < p_w < p_0$ where $p_{cr}$ is the critical pressure by $p_{cr} = p_0\left(\frac{2}{\gamma+1}\right)^{\frac{\gamma}{\gamma-1}}$, subsonic injection occurs. The parameters are obtained through isentropic relation:

$$p = p_w, \quad T = T_0\left(\frac{p}{p_0}\right)^{\frac{\gamma-1}{\gamma}}, \quad u = \sqrt{\frac{2\gamma}{\gamma-1} RT_0\left[1-\left(\frac{p}{p_0}\right)^{\frac{\gamma-1}{\gamma}}\right]}, \quad (11)$$

where $T_0$ is the total temperature, $\gamma$ is the specific heat ratio, and $R$ is gas constant of the mixture.

(3) For $p_w \leq p_{cr}$, sonic injection happens and

$$p = p_{cr}, \quad T = T_0\left(\frac{p}{p_0}\right)^{\frac{\gamma-1}{\gamma}}, \quad u = \sqrt{\frac{2\gamma}{\gamma+1}RT_0}\,. \tag{12}$$

The exit boundary is set as subsonic or supersonic outflow condition according to the local Mach number. For supersonic case, all variables are extrapolated from the interior field; for subsonic case, the pressure is set as the ambient pressure with the value 0.1 Mpa, and other variables are derived by extrapolation.

**2.5. Validating test**

The propagation of 1-D detonation wave in a tube is tested to validate the computation code and chemical reaction model. The tube is filled with stoichiometric hydrogen/air mixture at $p=2atm$ and $T=400K$. The tube is ignited by imposing high pressure ($30atm$) and temperature ($6000K$) at a small area $x \in [0, 0.5cm]$ on left closed-side. After a period of development, the detonation wave is formed and propagates towards the right open-end. In order to check possible influence of length-scale, three sets of grids with different intervals are tested, i.e. $\Delta x=0.1mm$, $0.3mm$ and $0.5mm$.

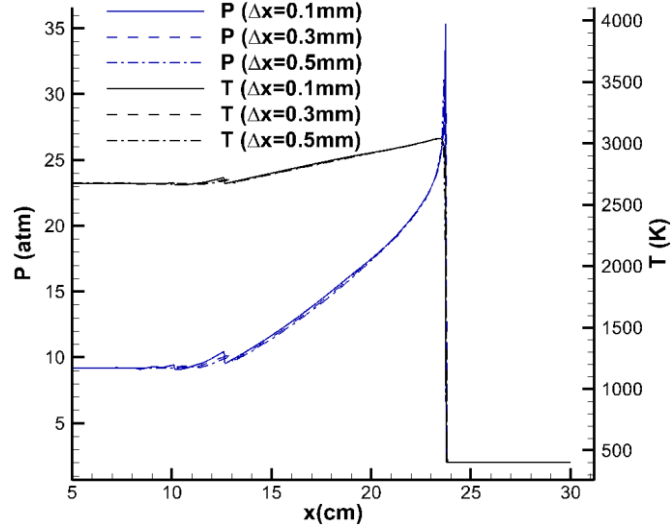

Fig. 2. Distributions of pressure and temperature at $t=120\mu s$ of 1-D detonation wave in different grids.

The parameter distributions at $t=120\mu s$ under $\Delta t=10^{-3}\mu s$ are shown in Fig. 2. The pressure and temperature discontinuity are matched together, which indicates computations in chosen grids can well capture the main physics of detonation. The predicted detonation properties at $t=120\mu s$, i.e. detonation velocity ($U_D$), pressure and temperature, are compared with that from Chapman-Jouguet (CJ) theory[9] as shown in Table 1. All errors are below 5%, which indicates an acceptable accuracy of the computation. Hence, the capability of predication on unsteady detonation is testified.

It is worth mentioning that the above test is also employed in Ref. [25], and more validating tests can be found there.

**Table 1. Comparison of properties of 1-D detonation.**

|  | $\Delta x$(mm) | $U_D$(m/s) | error | $P_{CJ}$(atm) | error | $T_{CJ}$(K) | error |
| --- | --- | --- | --- | --- | --- | --- | --- |
| Theoretical Value[9] | -- | 1979.1 | -- | 23.64 | -- | 3028.9 | -- |
| Computation | 0.1 | 1978.0 | 0.06% | 22.81 | 3.5% | 3024.5 | 0.15% |
|  | 0.3 | 1976.4 | 0.14% | 22.78 | 3.6% | 3023.5 | 0.18% |
|  | 0.5 | 1975.3 | 0.19% | 22.75 | 3.8% | 3022.9 | 0.20% |

## 3. Overall features of rotating detonation

For the 2-D chamber, the length in *y* direction is 10.0 *cm*, and the one in *x* direction is 4.0 *cm*. Stoichiometric hydrogen/air mixture is chosen as reactants. Total temperature $T_0$ and total pressure $p_0$ of the reservoir are 300K and 0.35MPa respectively. The initial conditions are set as follows: the top half of chamber at $y \in [2.0cm, 10.0cm]$ is filled with the premixture under the ambient temperature 300K and pressure 0.1MPa. The ignition is realized by imposing high temperature and pressure ($T = 3000K, P = 2Mpa$) on a small area at $x \in [0, 0.5cm]$ and $y \in [2.0cm, 2.5cm]$. Air is filled as the inert gas at the region $y \in [0, 2.0cm]$ behind the ignition area in order to make the detonation wave propagate in one direction only. Two sets of uniform grids are utilized to analyze the flowfield, i.e., relative coarse grid (*Grid1*) with the interval $\Delta = 0.2mm$ and fine grid (*Grid2*) with $\Delta = 0.133mm$.

After a few cycles, the detonation wave becomes quasi-stable and propagates from the bottom to the top. The results in two grids at typical instants are shown in Fig. 3. As shown by Fig. 3a, basic features reported in experiments [4] and simulations [8-19, 24] are reproduced by *Grid1*, which includes the detonation wave front A, the oblique shock wave B, the slip line C in the burnt gas between A and B, the combustible premixture layer D and the interface E between fresh premixture and products. Due to the continuous injection of fresh mixture into the chamber, a triangular layer of combustible mixture is formed, through which the detonation wave can sustain and propagate successively. With the increase of grids by using *Grid2*, unstable subtleties emerge, e.g. the appearance of vortical structures at the slip line as shown in Fig. 3b. In order to explore the mechanism, the analysis is necessary by way of approaches like linear instability analysis. Considering the large amplitude and unsteady nature of the instabilities, it is conventional to use the time-averaged flow field as the base flow, which is shown in Fig. 3c. After acquiring the knowledge from the analysis on the time-averaged result, the one in Fig. 3b can serve as the direct simulation of instabilities for comparison.

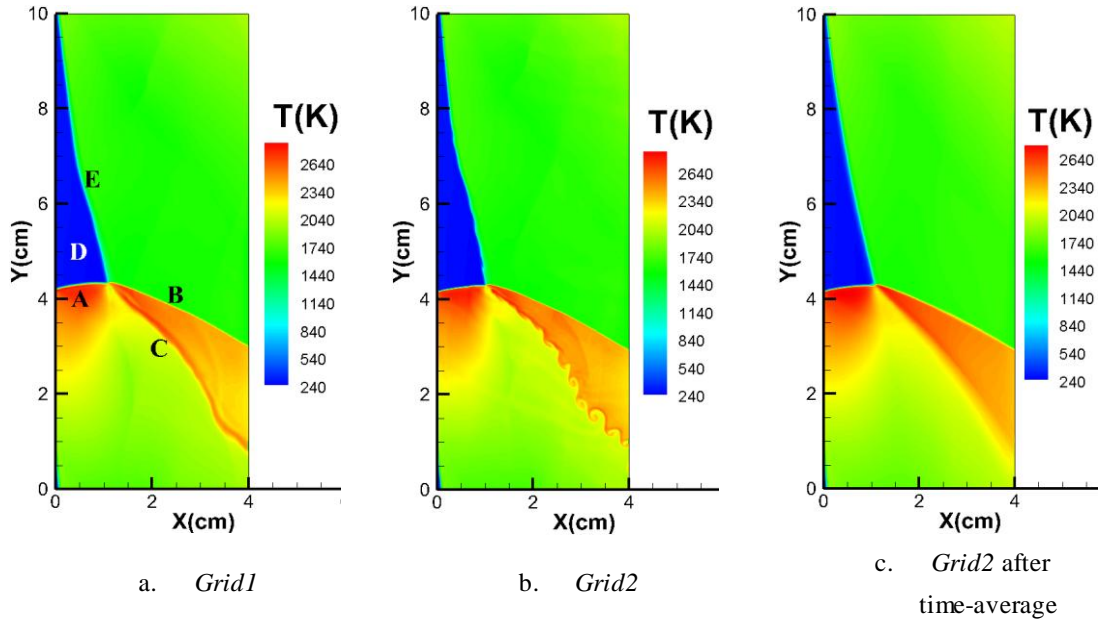

a. *Grid1*  b. *Grid2*  c. *Grid2* after time-average

Fig. 3. Temperature contours of 2-D rotating detonation in two sets of grid. In the result of *Grid1*, A-detonation wave, B-oblique shock wave, C-slip line, D-combustible mixture layer, E-the interface between fresh premixture and product.

To qualitatively check the operation of rotating detonation, the computation in *Grid2* is chosen for analysis, and similar results are obtained from that in *Grid1* as well. First, the law of mass conservation is checked by investigating the temporal variations of mass flux on inlet and exit in Fig. 4. From the figure, the mass fluxes oscillate around some fixed values and show statistically stable. The ratio between the inflow and outflow fluxes is about 1.0±0.05, which indicates the mass conservation is numerically established. The history of pressure at the point ($x$=0.4$cm$, $y$=8.0$cm$) near the inlet boundary is shown in Fig. 5. The good periodicity indicates the detonation wave propagates in a relative constant speed.

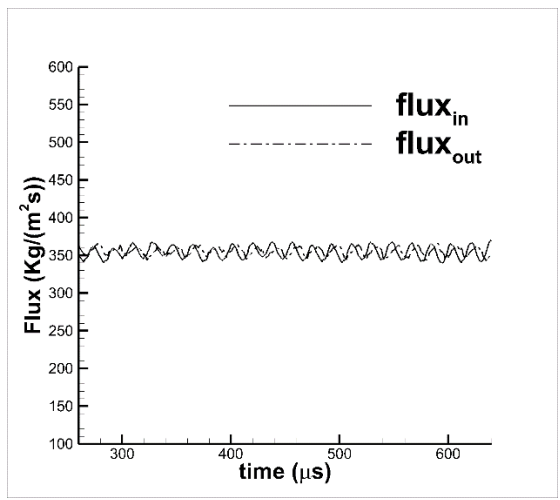 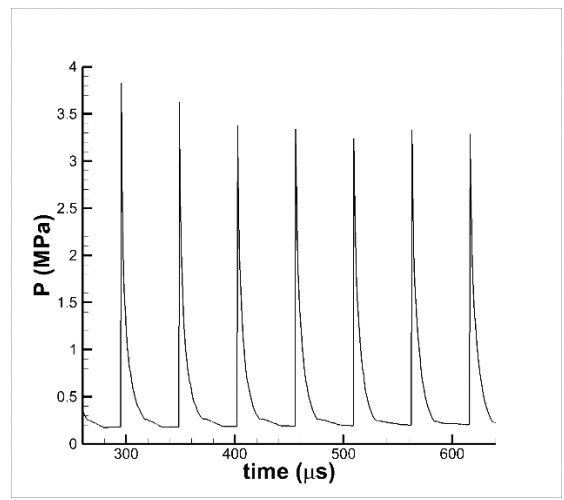

Fig. 4. Variation history of mass fluxes on inlet and exit in the case of *Grid2*.

Fig. 5. Variation history of pressure at the observation point ($x$=0.4$cm$, $y$=8.0$cm$) in the case of *Grid2*.

At last, some quantitative detonation properties in two cases are summarized in Table 2, where the inflow mass flow rate $\dot{m}_{in}$, frequency $f$, azimuthal speed $v_D$, and specific impulse $I_{sp}$ based on fuels are evaluated after taking average over several cycles (during $280\mu s$ -$600\mu s$). All results in two grids are quite close to each other and relative differences with respect to the results of *Grid*2 are within 1.0%. The good consistency implies that two grid resolutions have limited effect on referred detonation properties, although differences exist on the description of flow subtleties as shown in Fig. 3. Choosing the state of a typical point ahead of the detonation, the theoretical C-J velocity can be evaluated as 1996.4 *m/s*, and corresponding numerical difference of $v_D$ is about 7.3%. Hence current computations show well accuracy qualified for further analysis.

Table 2. Comparison of characteristics of 2-D detonation in two grids.

|  | $\dot{m}_{in}$ (Kg/($m^2$ s)) | $f$ (KHz) | $v_D$ (m/s) | $I_{sp}$ (/s) |
|---|---|---|---|---|
| *Grid1* | 352.1 | 18.81 | 1850.67 | 5197.9 |
| *Grid2* | 354.4 | 18.69 | 1849.22 | 5163.3 |
| Relative difference to *Grid2* | 0.6% | 0.6% | 0.08% | 0.7% |

## 4. Analysis of wake instability at the slip line

As mentioned in the introduction, the flow instability at the slip line has been discovered by simulations for a long time, where the notion of K-H mechanism was generally referred without enough analysis. Through carefully analysis in this study, an interpretation of wake instability is proposed, and discussions are made to clarify the following uncertainties: (1) What are the characteristics of the wake flow and the cause of its generation? (2) What are the unstable modes of wake instability and corresponding flow structures?

### 4.1. Characteristics and cause of wake generation

In order to facilitate discussion, a velocity transformation is introduced first, through which the detonation wave will keep approximately stationary in the new system. Specifically, the following operation is employed to derive a new azimuthal velocity $v'$ by

$$v' = v - v_D, \tag{13}$$

where $v$ is the original azimuthal velocity and $v_D$ can be found in Table. 2. As mentioned before, the time-averaged flow in *Grid2* is chosen as the base flow for instability analysis, and the velocity transformation is carried on it first.

After the velocity transformation, the injected reactants will flow toward the quasi-steady detonation wave, and the contour of velocity magnitude is shown in Fig. 6. In the figure, the arrow represents velocity vector at representative point for visualization. It can be seen that when the fresh mixture flows through the detonation wave, the velocity magnitude drops dramatically at first, but it recovers and increases quickly downstream. In the meanwhile, the velocity magnitude of old burnt products also decreases when they pass through the oblique shock. What is more, there is an obvious low speed zone which corresponds to the gas mixture passing through conjunction of the detonation and oblique shock wave. The velocity vector near the low speed zone is approximately parallel to the streamwise direction of the zone layout. In summary, a wake is generated from the joint of the detonation and oblique shock wave, which differs from the traditional contact discontinuity in mixing layers. It is worth mentioning that similar velocity deficit has been reported in Refs. [13, 19] by Mach number contours, but the possibility of different instability from the widely-believed K-H one is less aware of by literatures.

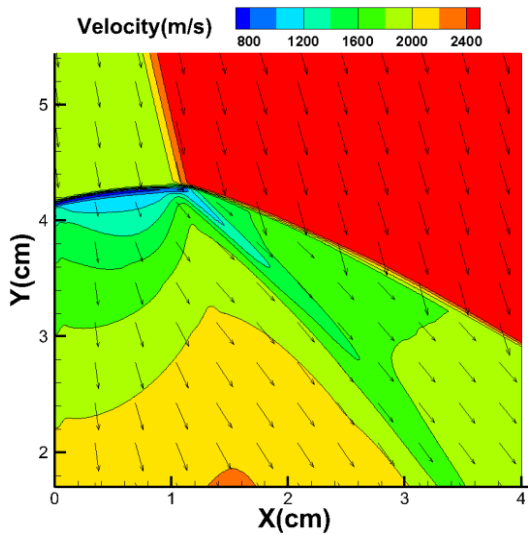 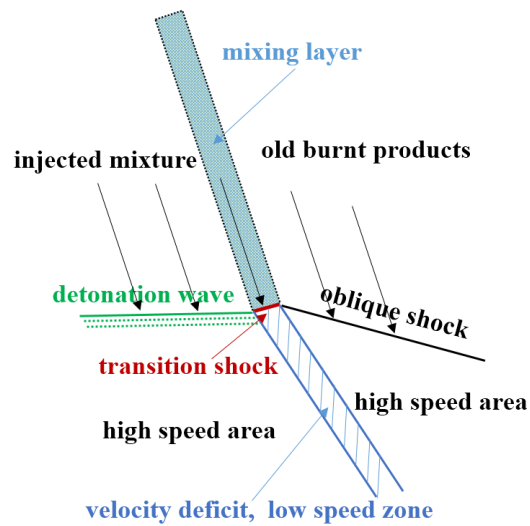

Fig. 6. Contours of velocity magnitude together with velocity vectors in the case of *Grid2* after time-average.

Fig. 7. Schematic on formation of velocity deficit.

Next, the cause of wake generation is analyzed. At least in Ref. [18], a schematic was proposed to describe the flow structures, where the oblique shock did not directly link to the detonation wave through a point, instead it shifted a little upstream and joined the lagged detonation wave through a small and slant stem (see Fig. 4 in the reference). Similar transition structure is also observed in current simulation as shown in Figs. 6 and 8a. Regarding the upstream shift of the oblique shock, Ref. [16] mentioned that the shock wave "propagates faster in the burnt product region than in the fresh mixture region". So it is conceivable that the existence of a tiny transition shock will be reasonable as the conjunction between the oblique shock and detonation wave. Under current velocity transformation, the newly burnt products of detonation have very high pressure, which drives the flow and increases its velocity rapidly during the expansion downstream. Hence, the overall velocity magnitude after the detonation wave becomes larger than that of the gas mixture passing through the transition shock. In the meanwhile, when the old products pass through the oblique shock, the flow decelerates, but its velocity magnitude is still rather large (the flow is found to be supersonic from the computation under current framework by Eq. (13)). As a result, a velocity deficit occurs along the slip line and the wake flow is generated. Through carefully checking, it is further found that the upstream flow before the transition shock comes from the mixing layer between the fresh reactants and old burnt products, therefore the velocity of which is relatively lower than that of the old burnt product and favors the generation of wake. Based on the understandings, a schematic model in Fig. 7 is proposed to illustrate the physics of wake generation. It is worth mentioning that the short stem in the figure is a simplification for the transition shock, and its rationality to explain the wake generation will be further illustrated next. In real complexity, the straight appearance of stem exists locally and the transition shock joins adjacent waves in smooth manner.

In order to check the rationality of the schematic in Fig. 7, numerical results are used for heuristic analysis. The motivation is that if the inflow condition and the geometry of the transition shock can be defined, the downstream properties could be predicted by shock relations and therefore compared with the computation. The derivation of inflow information and shock geometry is illustrated in Fig. 8a. In the figure, the white solid line is a streamline approximately passing through the center of the wake slot, and the state at *Point1* before the transition shock is used as the inflow condition. The outline of transition shock is approximately obtained by using the big jump of temperature gradient. The incident angle $\beta$ can be derived thereafter as $\beta \approx 73.76°$. Using the oblique shock relations with the absence of chemical reaction, predictions are made to compare with computations as shown in Figs. 8b-c. In the figure, the dashed line with subscript "1" denotes the state before transition shock at *Point1*, the dash-dot line with subscript "2" denotes predictions, and the red solid line represents numerical results on aforementioned streamline. From the figure, it can be seen that it is appropriate to choose *Point1* as the upstream location, and a reasonable agreement is achieved between the predictions and numerical results. Considering the absent consideration of chemical reactions, some differences noticed should be conceivable, e.g. the predicted temperature are higher than numerical counterparts.

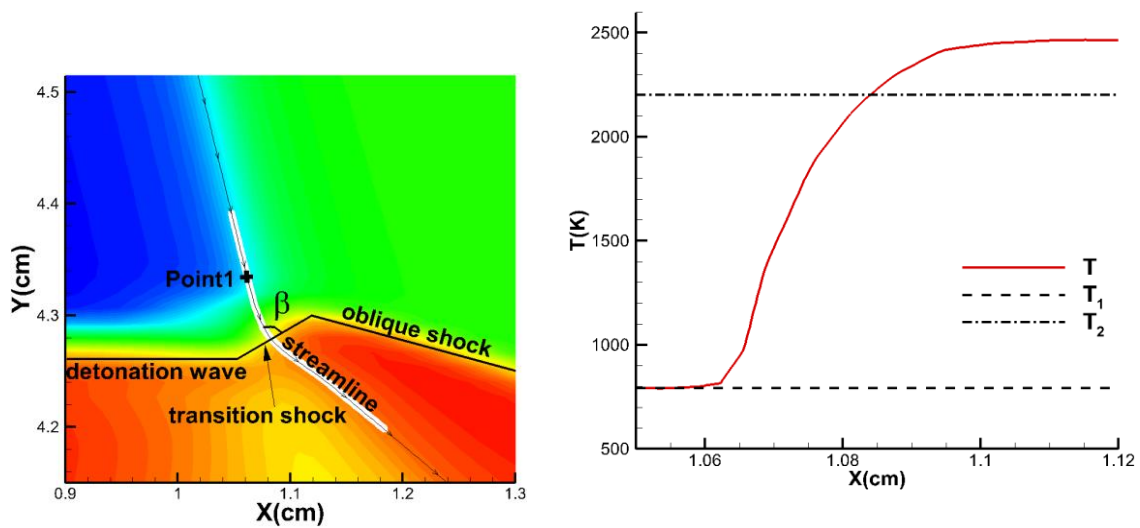

a.  Choose of streamline and its incident angle with regards to transition shock under temperature background

b.  Temperature

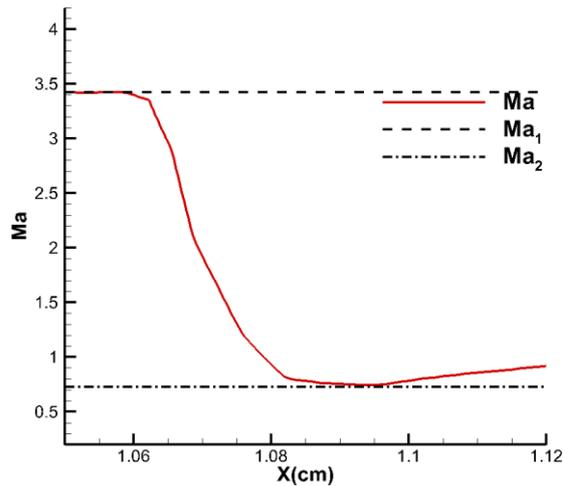

c.

Fig. 8. Parameter distributions at streamline passing through deficit center with post-shock references (with subscript "2") predicted from pre-shock states at *point1* (with subscript "1").

At last, in order to show the difference between the wake and canonical mixing layer, a discussion is made regarding vorticity distribution. It is well-known that the tangential-like velocity profile in mixing-layers results in a single vorticity layer, i.e., the vorticity has a concentrated distribution with one consistent direction in the mixing layer. For the wake velocity profile, two shears with opposite torques exist, which consequently yields two layers of vorticity with opposite directions. For demonstration, vorticity contours in *Grid2* are drawn in Fig. 9. From the time-averaged result, two vorticity layers with opposite directions along the slip line are shown in Fig. 9a, which are referred as double-layer of vorticity in this study. In more detail, the strength of positive vorticity is larger than that of negative one (see Fig. 11a). It is worth mentioning that in canonical 2-D mixing layer, analogous vortex pillows can be observed but their vorticity usually has only one direction as that in the original shear layer. In current situation, different structures evolve differently in two regions, namely, distinct vortex-like structures appear in the layer with positive vorticity, while the layer with negative vorticity only becomes curved and appear a relatively weak evolution(see Fig. 9b). The features indicate a specific wake which differs from the typical symmetric one after the cylinder, and the analysis will be made in the next section.

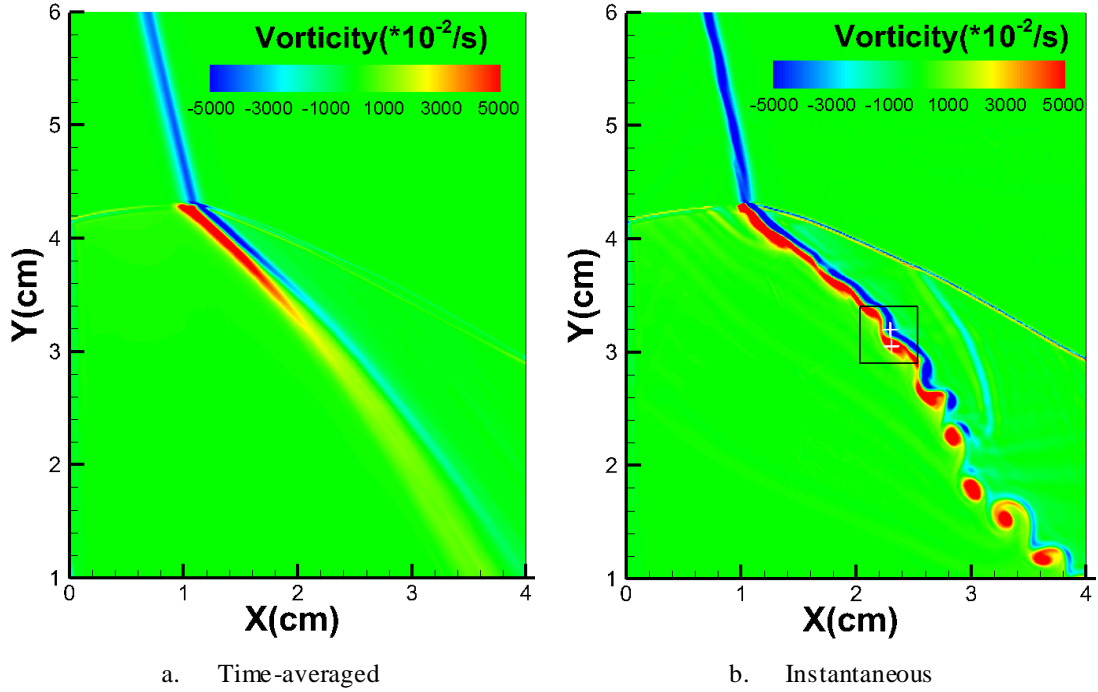

a.  Time-averaged        b.  Instantaneous

Fig. 9. Time-averaged and instantaneous vorticity contours in *Grid2*. In the instant field, two markers in zoomed-in window denotes initial positions for trace tracking.

### 4.2. Different unstable modes of wake instability and corresponding flow structures

Due to the wake profile of the velocity, the instability at the slip line in 2-D rotating detonation will be different from the canonical mixing layer. In order to facilitate analysis, a coordinate transformation is imposed on the transformed velocity field by Eq. (13), i.e., both coordinate translation and rotation are invoked, through which the origin is located at the start of the slip line, and the new horizontal axis $x_1$ will be placed along the slip line. After the transformation, the gas around $x_1$-axis in *Grid2* after time-average will flow approximately in parallel with the axis as shown in Fig. 10. Hence, the analysis by the linear stability theory (LST) based on the parallel-flow assumption can be employed.

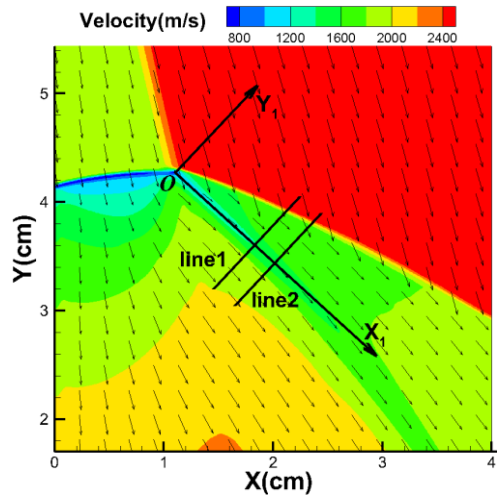

Fig. 10. Coordinate transformation and chosen lines for LST analysis in the case of *Grid2*

after time-average.

It is essential to choose proper parametric profiles as the base flow for instability analysis. After carefully checking the distributions at different locations, two representative lines, namely *line1* and *line2*, are chosen for analysis and shown in Fig. 10. The distributions of velocity $u_1$ and $v_1$ on these lines are shown in Fig. 11. In the figure, distributions are trimmed when $y_1$-coordinate is quite away from the slip line, because the parameters outside will either lose uniformity, violate the parallel-flow assumption or meet the oblique shock. Although distributions in two lines still show some inconformity, obvious similarity and shear feature are indicated. What is more, the distributions indicate a fairly good parallel feature of the flow, which is preferred by LST. Based on the above understanding, the two profiles in Fig. 11 are chosen in this study for LST analysis to provide heuristic knowledge on unstable mechanism.

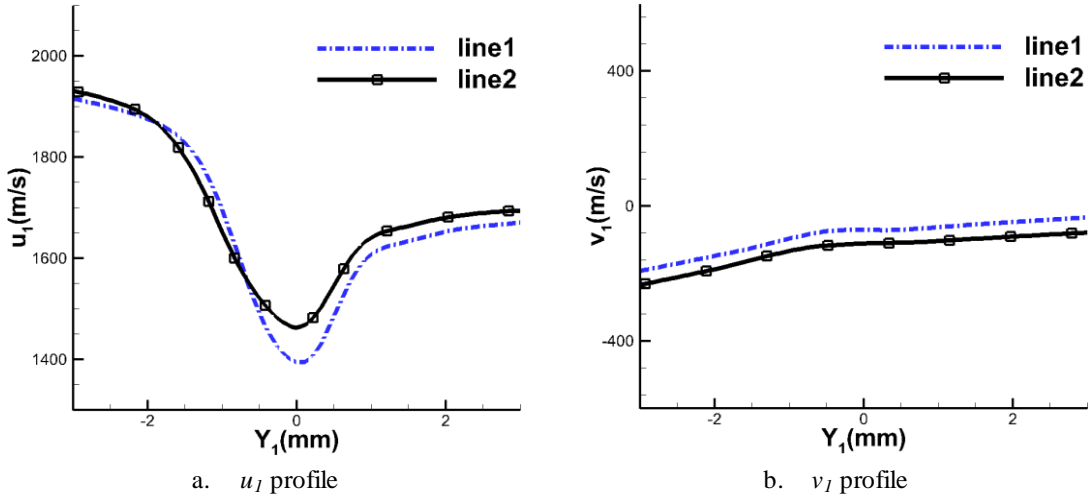

a.  $u_1$ profile      b.  $v_1$ profile

Fig. 11. Parameter distributions along *line1* and *line2* for LST analysis in the case of *Grid2*.

For each profile, the eigenvalue problem is solved to derive the solution in the travelling-wave form $\hat{\phi}(y_1)e^{i(\alpha x_1 - \omega t)}$, where $\alpha$ is the complex wave number, $\omega$ is the real disturbing frequency, and $\hat{\phi}(y_1)$ is the eigen-function of the variable $\phi(x_1, y_1, t)$. Through LST analysis, two branches of unstable modes are found. In each branch, the mode with maximum increase rate $(-\alpha_i)$ can be further specified (referred as A and B respectively), and corresponding results are summarized in Table 3. In the table, the phase speed $c$ is especially provided through $c = \omega/\alpha_r$. Comparing the two unstable modes regarding *line1*, mode B has higher frequency, shorter wave length, and especially higher increase rate. Same trends are also obtained at *line2*.

**Table 3. Characteristics of unstable modes by LST analysis based on parametric profiles on *line1* and *line2*.**

| Line | Unstable mode | $\omega$ (rad/s) | $\alpha_r$ (m$^{-1}$) | $-\alpha_i$ (m$^{-1}$) | $c$ (m/s) |
|---|---|---|---|---|---|
| line1 | A | $1.056 \times 10^6$ | $1.396 \times 10^3$ | 19.8 | 1512.98 |
| line1 | B | $0.731 \times 10^6$ | $0.880 \times 10^3$ | 48.1 | 1662.56 |
| line2 | A | $1.250 \times 10^6$ | $0.159 \times 10^3$ | 19.8 | 1568.21 |
| line2 | B | $0.622 \times 10^6$ | $0.724 \times 10^3$ | 35.2 | 1719.70 |

Given wave numbers and frequencies, eigen-functions $|\hat{u}_1|$ can be solved as well, and those of the velocity $u_1$ on *line1* are shown in Figs. 12a-b as examples to compare with numerical results. From the figure, it can be observed that the shape of $|\hat{u}_1|$ of mode A has a primary peak and its location is roughly at the region where $y_1$ is slightly larger than 0, while that of mode B has a double-peak appearance and its distribution is approximately symmetric to $y_1 \approx 0$. As shown by the wake profile in Fig. 11a, the positive vorticity layer has stronger velocity shear than that of the negative vorticity layer, therefore it is expected that the instability in the former layer behaves differently than that in the latter, which is consistent to the observation in Fig. 9b. The same analysis has also been made toward parametric distributions on *line2*, and similar results have been acquired as well. Considering the above outcomes of LST, it is conceivable that different unstable structures might occur in the layers after the transition shock.

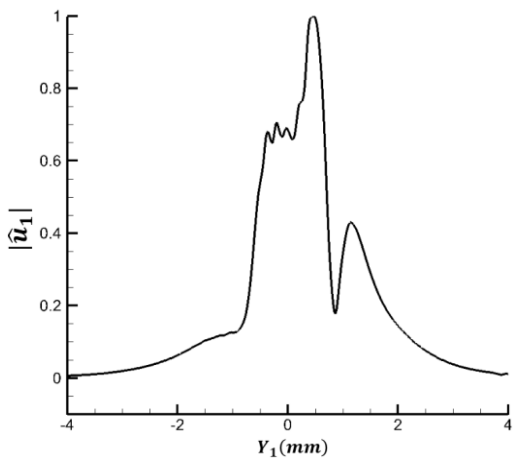 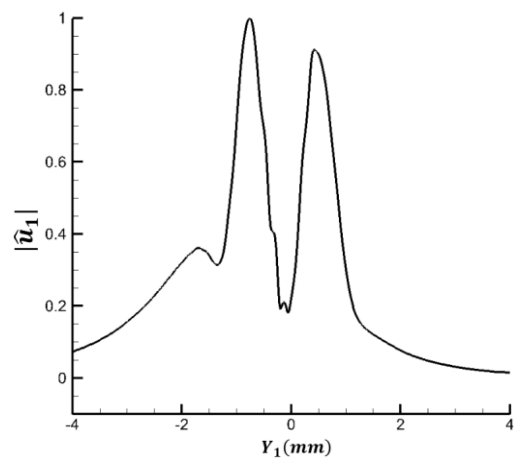

a. Eigen-functions of mode A by LST analysis   b. Eigen-functions of mode B by LST analysis

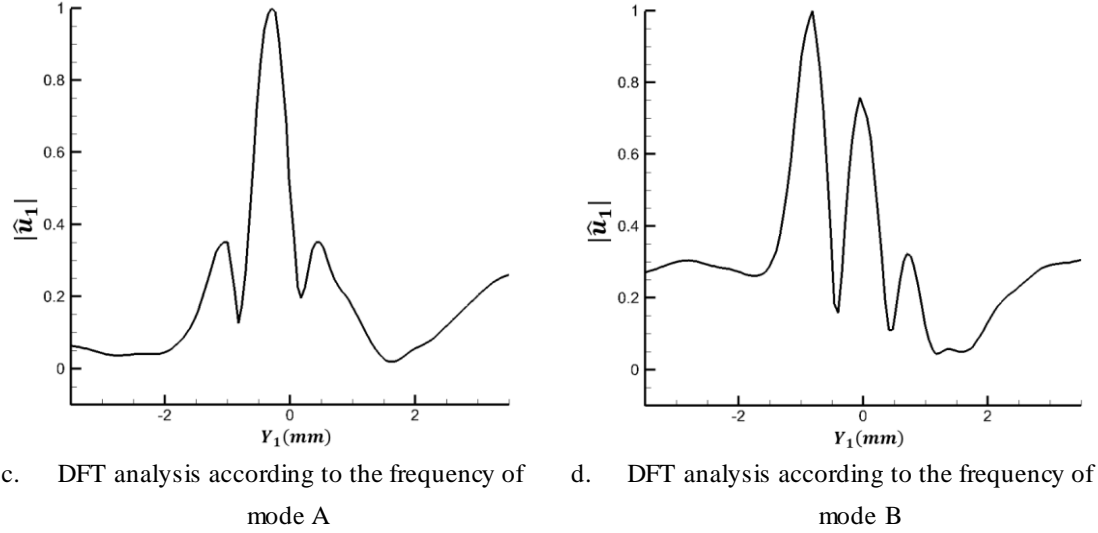

c. DFT analysis according to the frequency of mode A
d. DFT analysis according to the frequency of mode B

Fig. 12. Comparison of eigen-functions of $u_1$ along *line1* between predictions by LST and outcomes by DFT.

Based on above analysis, two comparisons are made between analytic predictions and numerical counterparts. In order to obtain the latter, oscillating parts of results in *Grid2* are first acquired by subtracting the averaged flow field from the instantaneous one, then discrete Fourier transformation (DFT) is carried out. The first comparison regards the eigen-function $\hat{u}_1$, where datum on *line1* are recorded for about one cycle of detonation operation and DFT is invoked afterwards. According to the frequencies of mode A and mode B in Table 3, the numerical presentation of $|\hat{u}_1|$ of two modes are derived by DFT and shown in Figs. 12c and 12d respectively. It is known from previous LST analysis that the mode A with the frequency $\omega=1.056\times10^6$ rad/s in Fig. 12a is so chosen that the mode would have the largest increase rate and therefore take the advantage over other disturbances. Thence, it is supposed that the mode A might take more effect in the formation of $\hat{u}_1$ in Fig. 12c, although the contributions of other modes could exist. Similar situations happen between results of Figs. 12b and 12d. It can be observed in Fig. 12c that the numerical $|\hat{u}_1|$ has a primary peak as Fig 12a, and in Fig. 12d the numerical $|\hat{u}_1|$ indicates an obvious double-peak structure similar to that in Fig. 12b. Theses consistencies are quite in favor of the validity of two unstable modes given by LST. In spite of the similarity, discrepancies are also observed. Firstly, secondary peaks appear around the primary ones in Fig. 12b and 12d. This might be attributed to the effects of other modes. Secondly, it is noticed that both sides of the numerical $|\hat{u}_1|$ do not show the attenuated distributions like the counterparts of LST. The possible reason is that in transformed velocity field by Eq. (13), the detonation wave, oblique shock and slip line are actually unsteady, therefore the investigated flow away from $x_1$ axis is oscillatory, which would yield non-damping distributions through DFT. In short, the resemblance of function profiles between LST and DFT indicates the two different unstable modes should take effect and contribute to the instability along the slip line.

The second comparison is about the velocity of the unstable structures of two modes. When an obvious unstable structure is generated, it is convenient to numerically track and derive its velocity. In this regard, two specific locations are chosen in positive and negative vorticity layers respectively, which are denoted by white crosses shown in Fig. 9b. After measuring the displacement $\Delta l$ of the tracking point for a short time interval $\Delta t$, the velocity of movement can be evaluated by $\Delta l / \Delta t$. In the meanwhile, the phase speed in Table 3 is used to approximately represent the speed of unstable structures for comparison. More specifically, because two sets of speed in Table 3 are provided based on parametric profiles on *line1* and *line2*, separate comparisons are provided in Table 4 accordingly. From the table, the difference is below 10.5%, which indicates a fair agreement between the analysis and computation.

**Table 4. Velocity comparison of unstable structures between analytic predictions and numerical results.**

| Layers | Velocity *(m/s)* | Difference regarding $c$ |
|---|---|---|
| Negative vorticity | 1404.00 | 7.2% (mode A of *line1*) |
| | | 10.5% (mode A of *line2*) |
| Positive vorticity | 1685.92 | 1.4% (mode B of *line1*) |
| | | 2.0% (mode B of *line2*) |

In summary, it can be understood that in the wake velocity profile, two kinds of unstable modes with different shape profiles and movement velocities are suggested to take effect, which would make the different layer of the wake evolve differently.

**5. Conclusions and discussions**

Base on Euler equations with the consideration of 7-species-and-8-reaction chemical model, the 2-D rotating detonation is simulated by using high order schemes. Two sets of grids are chosen in investigations, namely, the relative coarse one is used to generate a relative stable flow as the base flow for analysis, and the fine one is used to show nonlinear evolutions of instabilities. After making analysis and comparisons, the following conclusions are drawn:

1. The contact discontinuity along the slip line widely referred in literatures is not a mixing-layer-like structure, instead it is a wake of velocity. The cause of the wake generation is due to the transition shock between the detonation wave and oblique shock. Because of the wake profile, the vorticity distribution therein appears in a double-layer outlook, which is different from the canonical single layer in mixing-layers.

2. Numerical simulation indicates different evolutions exist in different vorticity layers. Based on the wake profile, LST analysis is made, and two main unstable modes are found which have different shape profiles and phase velocities. Numerical results are processed by DFT and similar structures are acquired as well. A gross coincidence is attained between the theoretical predictions and simulations, which favors the existence of two main unstable modes at the slip line.

It is worth mentioning that due to the inhomogeneity of density and pressure at the slip line, the effect of barolinic torque exists. Theoretically the torque would influence the evolution the vorticity of aforementioned two layers and exert influence on the instability. Further quantitative checking shows that the torque seems to be not strong and distributes locally. Therefore, according to the analysis in the paper, the wake stability is thought to bear an important role in the instability of the slip line, while other possible factors are still need to concern.

**Acknowledgements**

This work was sponsored by the National Science Foundation of China under the Grant number 91541105, and also partially supported by National Key Basic Research and Development 973 Program of China under Grant Number 2014CB744100. The second author is thankful to the discussion with Prof. Frank Lu.